\documentclass{acm_proc_article-sp}

\usepackage{graphicx}
\usepackage{epsfig}
\usepackage{amssymb}
\usepackage{amsmath}

\begin{document}

\title{Predicting Unemployment Claims Using Regional and Exogenous Signals: A Sparse Modeling Approach}
\subtitle{[Extended Abstract]}
%
%
%
%
%

\numberofauthors{3} 
%
\author{
%
%
\alignauthor
Avleen Bijral\\
       \affaddr{Microsoft Corporation}\\
       \affaddr{One Microsoft Way}\\
       \affaddr{Redmond, WA}\\
       \email{avbijral@microsoft.com}
\alignauthor
Richard Johnston\\
       \affaddr{Microsoft Corporation}\\
       \affaddr{One Microsoft Way}\\
       \affaddr{Redmond, WA}\\
       \email{rickyj@microsoft.com}
\alignauthor Juan Lavista Ferres\\
       \affaddr{Microsoft Corporation}\\
       \affaddr{One Microsoft Way}\\
       \affaddr{Redmond, WA}\\
       \email{jlavista@microsoft.com}
}


\maketitle
\begin{abstract}
In this paper we apply a time series based Vector Auto Regressive (VAR) approach to the problem of predicting unemployment insurance claims in different census regions of the United States.  Unemployment insurance claims data, reported weekly, are a leading indicator of the US unemployment rate.  Gathering weekly unemployment claims and aggregating by region, we model correlation between the different census regions. Additionally, we explore the use of external variables such as Bing search query volumes and URL site clicks related to unemployment claims.  To prevent any spurious predictors from appearing in the model we use sparse model based regularization. Preliminary results indicate that our approach is promising and in future work we plan to extend it to a larger set of predictors and a longer data range.
\end{abstract}


\section{Introduction}

\begin{figure}[!ht]
  \begin{center}
 \includegraphics[width=0.49\textwidth]{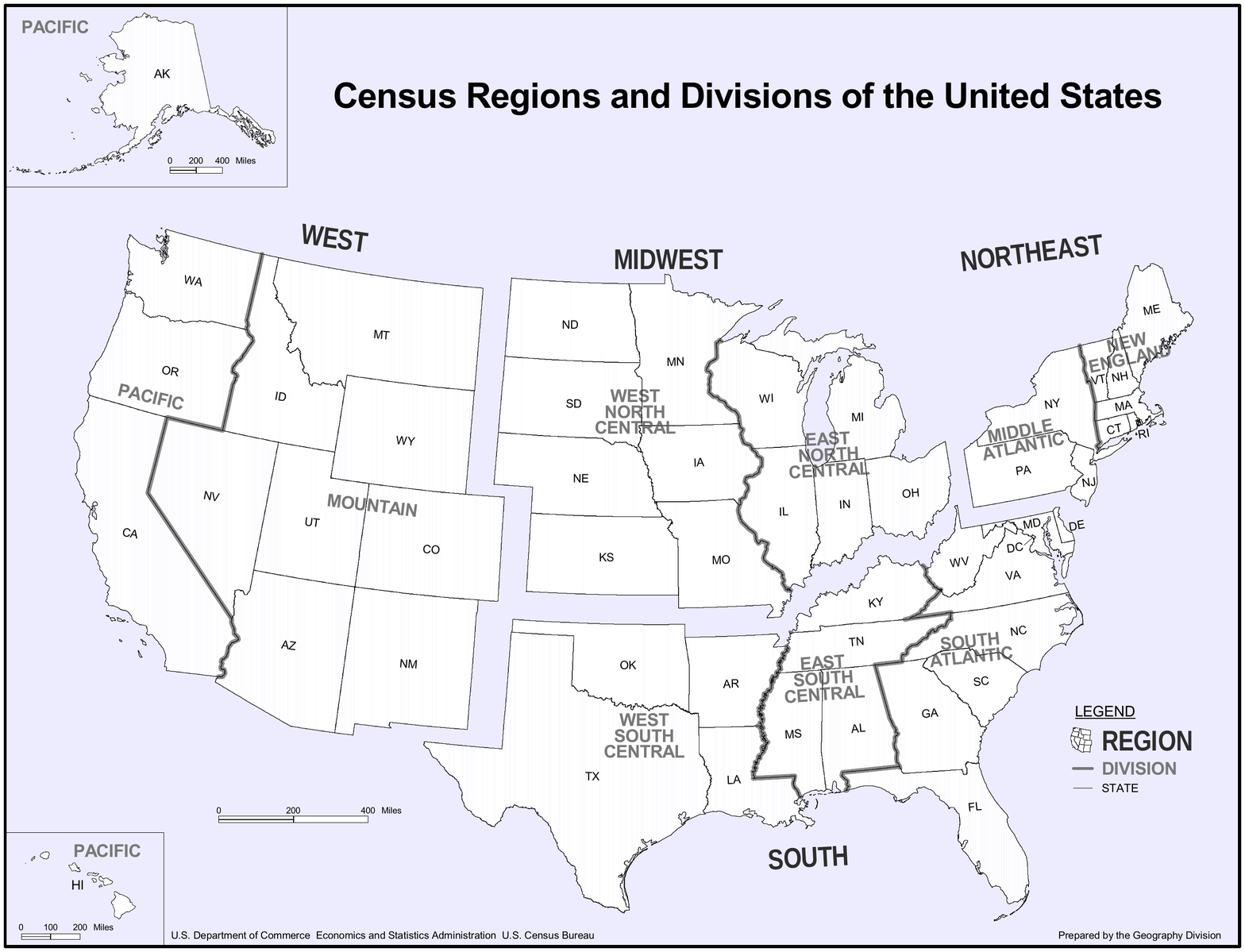}
 \end{center}
  \caption{Census regions in the United States.}
  \label{fig:censusRegions}
\end{figure}

\begin{figure}[!ht]
  \begin{center}
 \includegraphics[width=0.49\textwidth]{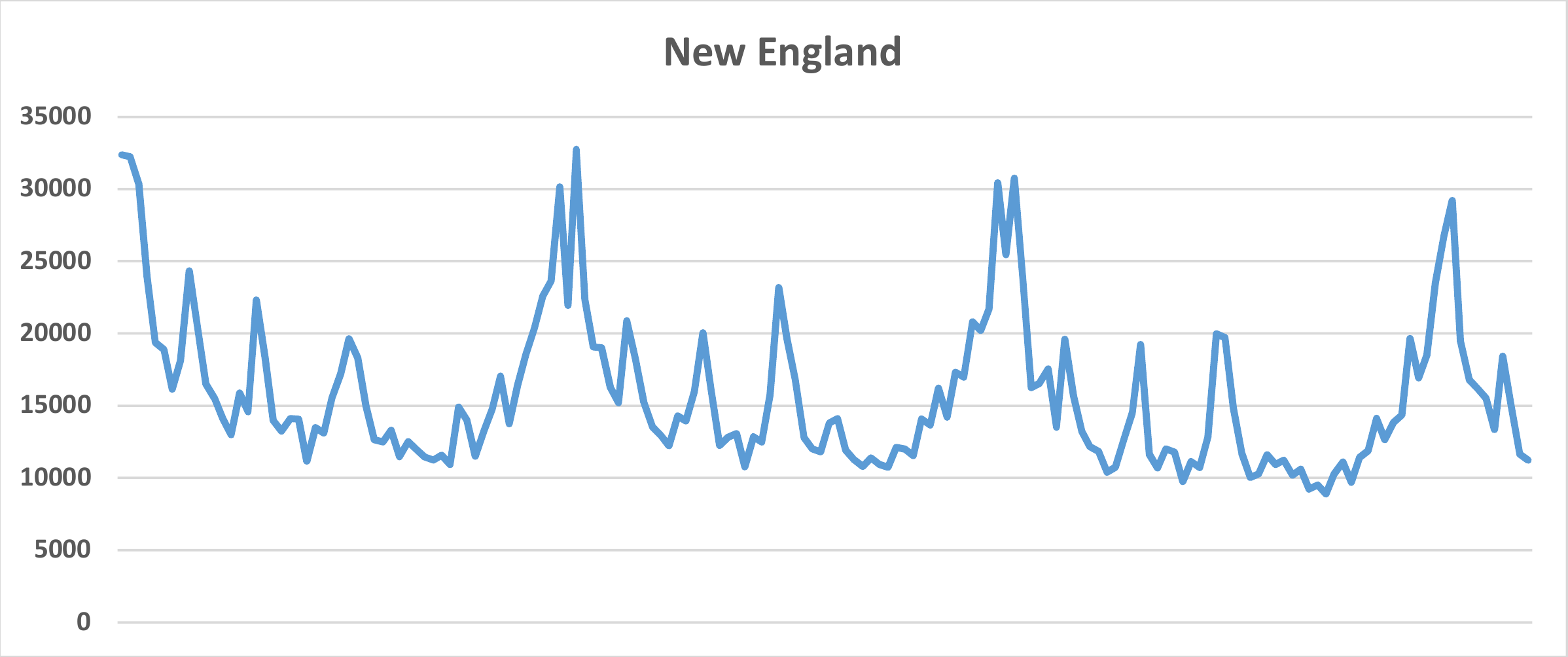}
 \end{center}
  \caption{New England - Weekly claims.}
  \label{fig:claimsRegions}
\end{figure}

The unemployment rate is a key indicator of the state of the economy, though in most places it is only reported monthly.  Unemployment insurance claims data (See Figure \eqref{fig:claimsRegions}), on the other hand, are reported weekly and are a leading indicator of the unemployment rate, making predicting the unemployment claims data very valuable.  Using historical unemployment claims data collected at the state level starting from January 2013 to April 2016, we aggregate by census region into nine separate regions shown in figure \eqref{fig:censusRegions}.  For each of these census regions we also gather Bing query term search volumes and URL click counts to use as external predictive variables.  A URL click is a click on the blue link that shows in search results and takes one to a new webpage.  Since unemployed, or soon to be unemployed individuals, likely search for information about unemployment and where to make claims, we hypothesize this information serves as a good concurrent and/or leading indicator of unemployment claims.  Searches on unemployment could signal general concern about losing ones job or a need for information post job loss.  URL clicks provide a way to essentially count how many people are going to each state's claims site, since the first search result is generally the source states respective unemployment site.  This data also shows cross region search activity.  Since insurance claims are made in the state one works in, not the state someone resides in,  states will show search activity for claims in their neighboring states.  Combining all of this cross information creates a relatively complex time series regression model, but provides an improvement over other simpler models making it worthwhile. Prior research using time series search data and unemployment does not analyze at this level of granularity nor develop a model with this degree of cross interactions between explanatory variables.    


\section{Notation and Model}

We propose a sparse vector auto regressive model with exogenous variables (VAR-X) for the weekly unemployment claims for $k=9$ census regions in the United States.  Let $\mathbf{Y}_t \in \Re^k$ be the vector of weekly unemployment claims and $\mathbf{X}_t \in \Re^m$ be the vector of exogenous signals consisting of weekly normalized query volumes and URL visits. Then the VAR-X model 
\begin{align}
\mathbf{Y}_t = \sum_{i=1}^p \boldsymbol{\Theta}^{(i)}\mathbf{Y}_{t-i} + \sum_{j=1}^s \boldsymbol{\beta}^{(i)}\mathbf{X}_{t-j} + \boldsymbol{\epsilon}_t
\label{eq:model}
\end{align}
where $\boldsymbol{\Theta}^{(i)} \in \Re^{k \times k}$ captures the regional dependence and $\boldsymbol{\beta}^{(i)} \in \Re^{k \times m}$ captures the cross-regional exogenous relationships. This model essentially captures the intuition that spatial proximity of the regions could indicate correlated claims. Additionally since we observed that people often search for unemployment claims in neighboring states, this indicates that model \eqref{eq:model} can also account for correlation between claims and exogenous signals in different regions.

The fundamental difference between our approach and that of \cite{} is that we infer the cross regional dependence by estimating the model using a sparsity inducing $\ell_1$-norm instead of imposing structural constraints. The optimization problem we solve is
\begin{align}
\min_{\boldsymbol{\beta},\boldsymbol{\Theta}} & \sum_{t=\max\{p,s\}+1}^T\left\|\mathbf{Y}_t - \sum_{i=1}^p \boldsymbol{\beta}^{(i)}\mathbf{Y}_{t-i} + \sum_{j=1}^s \boldsymbol{\beta}^{(i)}\mathbf{X}_{t-j}\right\|^2 \notag \\
&+ \lambda \left(\left\| \boldsymbol{\Theta} \right \|_1 + \left\| \boldsymbol{\beta} \right \|_1 \right) 
\end{align}
where $\boldsymbol{\Theta}$ and $\boldsymbol{\beta}$ are the stacked representations of the response and exogenous coefficient matrices.

\section{Related Work}

Choi and Varian (See \cite{Choi:Predicting}) used google trends data to predict initial jobless claims. Our work differs from theirs in the modeling technique, selection of predictor variables and level of data aggregation. They looked at jobless claims only at the national level.  Since initial jobless claims is often a leading indicator of the reported unemployment rate it is valuable to review works predicting unemployment rates too.  In this realm, Askitas and Zimmermann \cite{Askitas:Google} , Suhoy \cite{Query:Suhoy} and D'Amuri and Marcucci \cite{DAmuri:Google} use search data to predict unemployment in Germany, Israel, and the US respectively.  One downside of unemployment rate prediction is that most rates are reported on a monthly basis, while initial unemployment claims data are reported weekly.  Additionally, Choi and Varian do a review of predicting various variables in the present, including unemployment along with travel and consumer confidence.   We expand on this approach by using a more general forecasting model accounting for cross dependence between regions and also the different exogenous signals (query terms and clicks on links to specific sites to further improve predictions).

De Luna et al. \cite{DeLuna} looked at a Vector Auto Regressive (VAR) model for predicting monthly unemployment rates in the census regions. While this approach has some resemblance to our method, the important thing to note is that we do not force any spatial constraints in the model to infer any dependence between the regions. We achieve this goal by using a sparse penalty in the model. More importantly our model also incorporates exogenous signals. Search data provides an interesting source of boundary crossing information, since an individual can work in one state or region but live in another where their search occurs.

\section{Data and Estimation}

Unemployment insurance claims data (or initial jobs claims data) is not released until 8:30 A.M. Thursday on the week following the claim. E.g., the week ending April 9, 2016 is not publicly available from the Office of Unemployment Insurance, Employment and Training until Thursday April 14, 2016. Using Bing data we can obtain search related data in almost real time, providing information in advance of public announcement dates. Moreover we can use this data to provide predictions two weeks ahead for a more accurate assessment of a leading indicator of the economy in the near future.  We believe this is possible because individuals likely start searching unemployment related terms before they actually file for a claim.

The unemployment claims based weekly data reports are archived and accessible at the state level from the Federal Reserve Bank of St. Louis Economic data. We take this data and aggregate at the census region level.  The exogenous data is derived from historical searches and URL clicks.  We look specifically at the queries containing the term 'unemployment benefits' and related suggested query terms including 'file for', 'sign up for' and 'apply for' plus 'unemployment' and 'check unemployment status.'  In addition, we include the term 'unemployment' plus each state name in every state.  Searches for each state name in every region allows us to identify cross region relevant searches.  For example, an individual who lives in Alabama may work in Florida.  This person might search for unemployment at home (in Alabama) but make claims in Florida.  Additionally, we include click counts on the top URL results for these terms.  Including click counts of the top URL is valuable, especially in the state case because the majority of the time the first search result link is the respective state's unemployment insurance site (each state has its own unique site).  The region (and state) of origin for each query and click is found using a reverse IP look up.  Aggregation at the regional level is required to reduce the noise from sparse data occurring at the state level in search and URL results allowing for the inclusion of more terms.     

Finally, the query volume for each search time is normalized by considering the log transformed ratio of relevant query volume and the total number of distinct searches for that week. We currently only look at $178$ weeks of data and we believe that with a longer history we could achieve even better prediction rates. Additionally, we are currently working on using more exogenous signals to improve the prediction accuracy.

As we can see in Figure \eqref{fig:claimsRegions} there is a seasonal component ($52$ weeks) to the claims data we seasonal difference before we estimate the model. After differencing we use $1/3$rd  of the data for training the model, $1/3$rd for validation and the rest for testing. We use one-step rolling prediction to choose the regularization parameter in the cross validation procedure. The forecast on the test period is a one-step ahead rolling prediction. Note that that in our case we have more exogenous signals (and hence parameters) than data points but the use of regularization allows us to estimate a model even in this data deficient case.

Because the URL visits and the query volume can be sparse, we created two exogenous series for every region, one consisting of the average over all query volumes for that region and another for the average URL clicks for the region. The average query volumes are normalized by the total volume for that week for every region. This is expected to smooth out noise as well as handle missing data issues. Thus we have $9$ claims time series and $18$ exogenous series and using the VAR-X approach we hope to be able to incorporate any correlations across regions (both in the exogenous and the response). 

We compare the performance of the VAR-X models with the following time series
\begin{enumerate}
\item (A)-URL exogenous only (VAR-X)
\item (B)-Query exogenous only (VAR-X)
\item (C)-All (VAR-X)
\item (D)-No Exogenous (VAR Only)
\end{enumerate}

To estimate the model we used the optimization techniques developed in \cite{bigVAR} and the evaluation metric is the root mean squared error for the test period. For the VAR-X models, $p=2$ and $s=1$ gave the best performance. This supports the intuition that there is a lag of $1$-week between claim relevant searches and the actual claim application. 

\section{Empirical Results}

\begin{figure}[!ht]
  \begin{center}
 \includegraphics[width=0.5\textwidth]{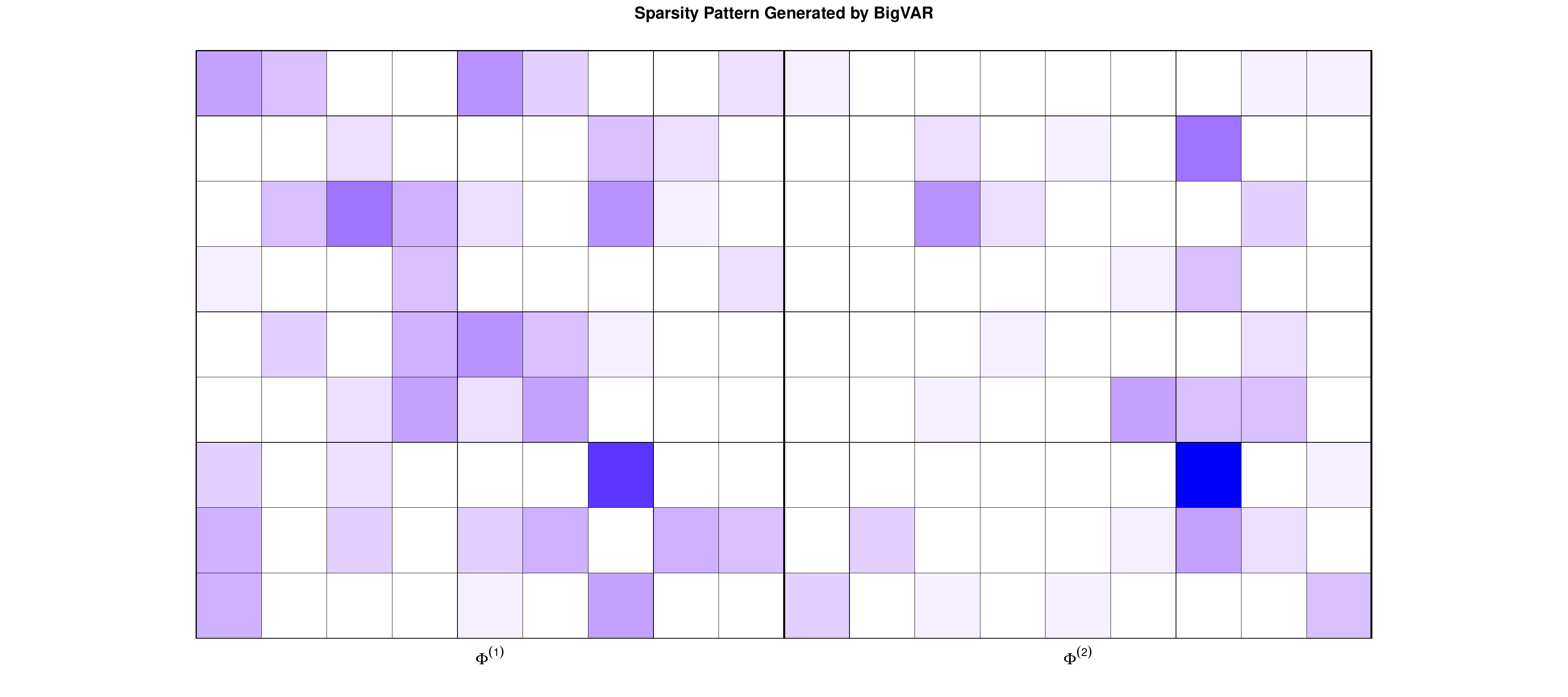}
 \includegraphics[width=0.5\textwidth]{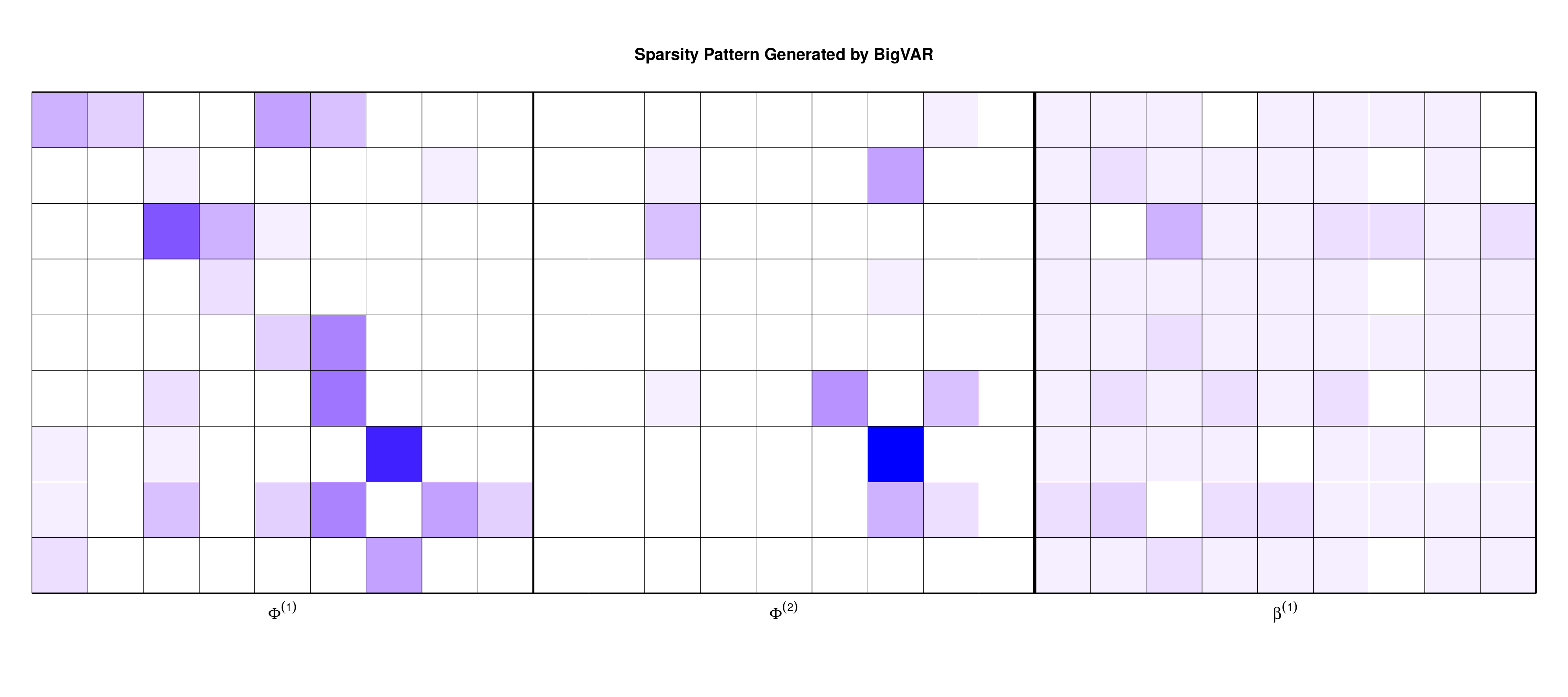}
 \end{center}
  \caption{a) Predictive Regional Dependence - The Sparsity pattern estimated reveals the cross region correlation from left to right -  Mid-Atlantic, New England, East North Central, West North Central, West South Central, East South Central, Mountain, Pacific and South Atlantic.
  b) Sparsity pattern with the exogenous queries included. We observe that the queries in a few regions have nonzero significant values and might possible help reduce prediction error.}
  \label{fig:sparsityPlot}
\end{figure}

\begin{table}
\centering
\begin{tabular}{|c|c|c|c|c|} \hline
(Region) & (A) & (B)& (C) & (D)\\ \hline
Mid-Atlantic & 0.06 & 0.07  & 0.07 & 0.06\\ \hline
New England& 0.08 & 0.08 & 0.09 & 0.08\\ \hline
East North Central & 0.12 & 0.123 & \textbf{0.10} & 0.13\\ \hline
West North Central & 0.05 & 0.047 & 0.048 & 0.046  \\ \hline
West South Central & 0.11 & 0.12 & 0.12 &  0.11  \\ \hline
East South Central & 0.153 & 0.155 & 0.16 & 0.15 \\ \hline
Mountain & 0.072 & 0.07 & 0.069 & 0.07 \\ \hline
Pacific & \textbf{0.14} & 0.15 & \textbf{0.14} & 0.16 \\ \hline
South Atlantic & 0.09 & 0.082 & 0.09 & 0.07 \\ \hline
\hline\end{tabular}
\caption{RMSE on the test weeks using different signals. Results indicate that exogenous signals can help in a few regions. In the others a pure VAR based approach works just as well, indicating sometimes the historical signal in a VAR model is sufficiently powerful.}
\label{tab:rmse}
\end{table}

Figure \eqref{fig:sparsityPlot} displays the estimated coefficient matrices $9 \times 9$ for a pure VAR model estimated by the sparse VAR model. The darker the entry indicates a larger magnitude of the estimated coefficient. The figure reveals a cross-region dependence especially at lag $1$. One can observe that there is some geographic dependence in the pattern of non-zeroes. Thus the sparse framework can capture these dependencies without encoding any prior information as is done in \cite{DeLuna}. Moreover there could be other similarities between regions that are either unknown or difficult to encode. We believe that the sparse modeling approach can help us identify these latent relationships using only the corresponding predictive power on a hold out set. Moreover when we include the exogenous signals the dependence on the history can be partly compensated, as indicated by the lighter sparsity pattern in the auto-regressive coefficient matrices. 
\begin{figure}[!ht]
  \begin{center}
 \includegraphics[width=0.49\textwidth]{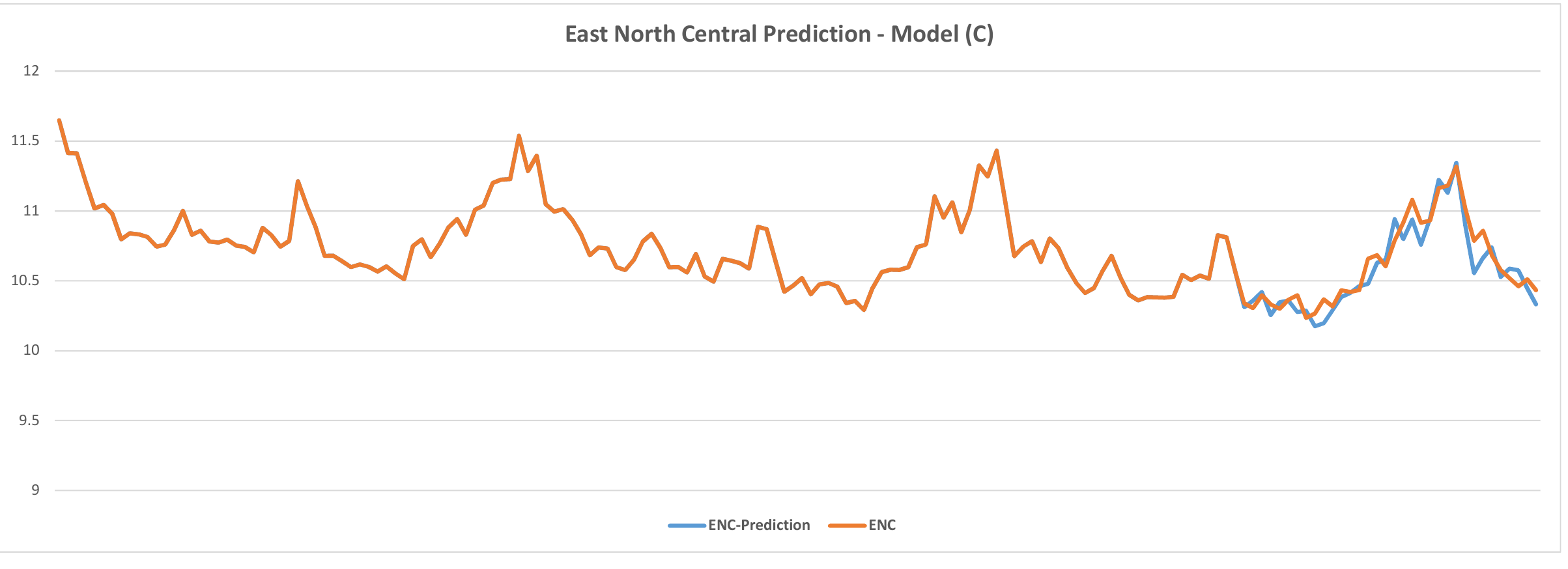}
 \end{center}
  \caption{East North Central Predictions vs Actuals}
  \label{fig:ENCPreds}
\end{figure}

Table \eqref{tab:rmse} shows the RMSE for the different regions on the test periods. The results indicate that except for a few cases (in bold) the VAR model with no exogenous variables is itself sufficient and can sometimes be supplemented with exogenous signals to improve prediction accuracy. A sample prediction using Model (C) for the East North Central region is shown in Figure \eqref{fig:ENCPreds}. We are currently trying to improve the accuracy in some regions using a more principled approach to find predictive URLs and queries. However our preliminary results do indicate that there is merit in using model such as \eqref{eq:model}. 

\section{Conclusions}
In this paper we proposed a model to predict weekly unemployment claims, which is a leading indicator of the unemployment rate. The VAR-X model incorporates predictive dependence between the different census regions and includes dependence on exogenous signals in the form of normalized URL and query volume. This data was collected for a period of $178$ weeks by looking at a fixed set of queries and the set of first URLs clicked after entering the query.

Our preliminary results indicate that except for a few regions where the exogenous signals seem to help improve the prediction accuracy a little, the cross-region dependence appears to be very useful. We hypothesize that this could be due to not just spatial proximity but also due to socio-economic similarities between the census regions.

We are currently exploring the use of queries and URL visits in a more systematic fashion, where we start with a large pool of queries and URLs and then filter out the predictive ones based on our sparse model and a holdout period. We believe that with data going back to 2011 and a much larger set of query/URL candidates we can can achieve better prediction accuracy.

\section{Acknowledgments}
The authors would like to acknowledge the help of the Bing team in support of collecting and gathering the needed search log data.

%
\bibliographystyle{abbrv}
\bibliography{sigproc}  

\begin{thebibliography}{1}

\bibitem{Askitas:Google}
N.~Askitas and K.~F. Zimmermann.
\newblock Google econometrics and unemployment forecasting.
\newblock {\em Technical report, SSRN}, 899, 2010.

\bibitem{Choi:Predicting}
H.~Choi and H.~Varian.
\newblock Predicting the present with google trends.
\newblock 2011.

\bibitem{DAmuri:Google}
F.~D'Amuri and J.~Marcucci.
\newblock Google it! forecasting the us unemployment rate with a google job
  search index.
\newblock {\em SSRN}, 2010.

\bibitem{DeLuna}
X.~De~Luna and M.~G. Genton.
\newblock Spatio-temporal autoregressive models for u.s. unemployment rate.
\newblock {\em Advances in Econometrics}, Volume 18 - Spatial and
  Spatiotemporal Econometrics 2004.

\bibitem{bigVAR}
W.~Nicholson and J.~Bien.
\newblock Varx-l: Structured regularization for large vector autoregressions
  with exogenous variables.
\newblock 2015.

\bibitem{Query:Suhoy}
T.~Suhoy.
\newblock Query indices and a 2008 downturn: Israeli data.
\newblock {\em Technical report, Bank of Israel}, 2009.

\end{thebibliography}
%
%

\end{document}